\documentclass{aa}
\usepackage{graphicx}
\usepackage{rotating}
\def\gtrsim{\mathrel{\hbox{\rlap{\hbox{\lower4pt\hbox{$\sim$}}}\hbox{$>$}}}}
\def\ltsim{\mathrel{\hbox{\rlap{\hbox{\lower4pt\hbox{$\sim$}}}\hbox{$<$}}}}

\begin{document}

\title{Variations of the He\,II $\lambda$1640\ Line in B0e--B2.5e Stars }
\titlerunning{He\,II  $\lambda$1640\ line variations in Be stars }
\author{M.A. Smith\inst{1}}
\authorrunning {Smith }
\institute{Science Programs, Computer Sciences Corporation,
Space Telescope Science Institute, 3700 San Martin Dr., 
Baltimore, MD 21218 ~~Email:~ msmith@stsci.edu 
 }

   \date{Received ??; accepted ??}

\abstract{Using the  {\it International Ultraviolet Explorer} data archive,
we have examined the {\it SWP} echellograms of 74 B0--B2.5e stars for
statistically significant fluctuations in the He\,II (``H$\alpha$") 
$\lambda$1640 line profile. In this sample we found 
that the He\,II line is occasionally variable in 10 stars over short
to long timescales. The He\,II-variable stars discovered are
$\lambda$\,Eri, $\omega$\,Ori, $\mu$\,Cen, 6\,Cep, HD\,67536, 
$\psi^1$\,Ori, $\eta$\,Cen, $\pi$\,Aqr, 2\,Vul, and
19\,Mon.  The most frequent two types of variability are an extended
blue wing absorption and a weakening of the line along the profile.
Other types of variability are a weak emission in the red wing
and occasionally a narrow emission feature. In the overwhelming number of 
cases, the C\,IV resonance doublet exhibits a similar response; rarely, it
can exhibit a variation in the opposite sense. 
Similar responses are also often seen in the Si\,IV doublet,
and occasionally even the Si\,III $\lambda$1206 line.  We interpret the
weakenings of He\,II and of high-velocity absorptions of C\,IV to 
localized decreases in the photospheric temperature, although this may not 
be a unique interpretation. We discuss the variable blue wing absorptions 
and red wing emissions in terms of changes in the velocity law and mass 
flux carried by the wind. In the latter case, recent experimental models by
Venero, Cidale, \& Ringuelet require that during such events the wind
must be heated by 35,000\,K at some distance from the star.
}
\maketitle

%\newpage  
\section{Introduction}
\label{int}

 Together with the hydrogen lines, the lines of helium are among the
most important in the spectra of hot stars. In atmospheres of B stars and
most O stars, the dominant ion stage of helium is He$^+$, and the strongest
of the He\,II features is the $\lambda$1640 (``H$\alpha$") line.
In O stars this line is formed substantially in the wind -- so much
so in supergiants that the line generally develops a strong P\,Cygni
emission structure. For spectral types later than O8--09, the line 
decreases in strength, but it remains visible as a photospheric diagnostic
for spectra as late as B2.5 (Peters 1990, Rountree \& Sonneborn 1991). 

  The He\,II $\lambda$1640 line is actually a complex of seven permitted
transitions arising from lower levels at 40.8 eV. Its effective centroid
wavelength is 1040.42\,\AA.~ In the outer atmospheres of hot stars
this line is formed by the
photoionization of He$^{2+}$ by extreme UV ($<$$\lambda$228) photons,
followed by recombination. The density and temperature sensitivities
insures that the line's formation occurs substantially in 
the base of the wind or within the photosphere
for O and B stars near the main sequence, respectively.

  The $\lambda$1640 line is mildly sensitive to departures from LTE in the
He$^{1+}$ atom. As a result, the strengths computed from non-LTE models tend
to be slightly stronger than those from LTE models. 
%(e.g., Smith, Sterken, \& Fullerton 2005). 
Since these effects are relatively small,
there appears to be no major difficulties fitting this line approximately
with conventional blanketed non-LTE model atmospheres.  Auer \& Mihalas (1972)
suggested that the near coincidence of central wavelengths of the He\,II
and some hydrogen lines could enhance emission of He\,II $\lambda$4686
and $\lambda$1640 through optical pumping (Bowen fluorescence).  However,
using more recent atomic parameters, Herrero (1987) demonstrated that
these effects are negligible.  Recently,  Venero, Cidale, \& Ringulet (2000;
``VCR") have considered the behavior of the $\lambda$1640 line
for model atmospheres with $T_{\rm eff}$ = 25,000\,K and strong, isotropic,
and heated winds. These authors find that even for model
atmospheres of early-type B stars a fast and/or heated wind can alter the
underlying photospheric profile. For example, in these models $\lambda$1640
undergoes a near maximum
{\it absorption} strength in winds having a temperature determined by
radiative equilibrium, i.e., with $T_o$/$T_{\rm eff}$ = 0.6--0.8. However,
if the wind is heated to 10,000\,K above the $T_{\rm eff}$, then emissions
will be produced in one or both of the line wings. Thus, isotropic, heated
($T_o$ $\ge$ $T_{\rm eff}$) winds produce a P\,Cygni-type profile, that is,
with a distinctly blueshifted absorption and slightly redshifted emission.
For standard wind models for early-type Be stars
(unheated winds, with \.M $\sim$ 1$\times$10$^{-8}$  M$_{\odot}$ yr$^{-1}$),
emission should be absent or undetectable. 
Even in the spectrum of the O9\,V star 10\,Lac, with its
mass loss rate of 1.7$\times$10$^{-7}$ M$_{\odot}$ yr$^{-1}$, the $\lambda$1640
profile is unshifted and has no redshifted emission (see Figure\,18a of
Brandt et al. 1998). It can be added that unpublished thesis work by
Cidale (1993) suggests that these same trends continue with emission in 
C\,IV. As a postulated hot temperature rise in these models is moved 
outward, a mild red emission component in C\,IV is enhanced while 
the absorption component remains almost the same. Some of our examples 
of observed variations below will illustrate this behavior.

  The $\lambda$1640 line has been interpreted by some authors to be stable in
strength and thus to be a good measure of an O or early B star's effective
temperature.  As detailed below, variability in this line has only seldom
been reported in hot, chemically homogeneous, single stars. 
According to the VCR models, any variability of this line
would require substantial changes in the star's effective temperature,
mass loss rate, or restructuring of the wind stratification. In cool
stars $\lambda$1640 variations  arise from
variable EUV irradiation in their chromospheres (Linsky et al. 1998). 
In this paper we will test several claims in the literature for
$\lambda$1640 variability in Be star spectra and extend the search
for this variability to a larger sample of B stars. Positive
results of this search will be placed in the general context of the
predictions of the VCR's {\it ad hoc} and nonstandard models of winds in
hot stars.

\section{Historical Variations of He\,II $\lambda$1640 in Be Stars }
\label{obsvar}

  VCR's models predict that $\lambda$1640 variations are caused
by dramatic changes in an O or B star's wind structure. This is
consistent with the report by Peters (1990) that a weak correlation exists 
between the strength of the $\lambda$1640 line and the wind component of CIV
$\lambda$1550i of $\lambda$\,Eri (B2e). Peters noted that these variations 
depend in part on where the star is in its ``wind oscillation cycle."

   A second report of correlations between variations of He\,II and another 
line in a B star came from a campaign with a ground-based telescope
and the {\it International Ultraviolet Explorer} ({\it IUE}) during three
8-hour ``shifts" on $\lambda$\,Eri. According to Smith
et al. (1996), decreases in the $\lambda$1640 and C\,IV resonance doublet
absorptions coincided with the creation of a ``dimple" on five occasions on
1990 October 21 and October 22 (see also Smith \& Polidan 1993, Smith et al.
1996) in the line profile of  $\lambda$6678 and other optical He\,I
lines.\footnote{Dimples are features appearing anywhere on the line
profile (so far only for He\,I lines), described as central absorptions
flanked by weak quasi-emission
ons on either side that largely compensate the central absorption. These
features typically have a lifetime of 2--3 hours. Smith et al. (1996) reported
that they may occur in the $\lambda$6678 line of a few other Be stars.}
The timescale for these changes was as short as ${\frac 12}$ hour.
These line profile transients hint at the presence of rapid, possibly
magnetic, activity close to the surface of the star. 
A connection with $\lambda$1640 changes would
further tie this activity to the photosphere.  In view of this short
history, we began our search for $\lambda$1640 activity in spectra
of early-type Be stars for which claims have been made preferentially for 
the presence of magnetic fields. Because the theoretical predictions are
that changes in $\lambda$1640 strength should be found in the dense, 
rapidly accelerating regions of winds of Be stars. It is logical to search
the spectra of stars which have known histories of variable wind components
of UV resonance lines.

\section{Observations and Data Analysis}

\subsection{Reduction of IUE Data}

  The ultraviolet  data for these programs are extant high-dispersion 
{\it IUE} echellograms obtained through the large aperture of the Short 
Wavelength Prime (SWP) camera. These data were obtained from the MAST
archive.\footnote{Multi-Mission Archive at Space Telescope Science 
Institute, in contract to the National Aeronautics and Space
Administration.} An IDL program was written
to read all spectra obtained for a star in the orders containing the
echelle orders of $\lambda$1640 and the resonance lines of C\,IV, N\,V, 
Si\,III, and Si\,IV. The spectra were cross correlated against one another
to place them on a common wavelength scale an co-added. Small rescalings
were then applied to the individual spectra to force their continua to the 
same level. Generally, these orders contain imprints of the
instrumental calibration ``reseaux" etched
onto the faceplates of the camera. The flux dips they cause are omitted in 
our plots to avoid possible confusion.

\subsection{Analysis of the $\lambda$1640 spectral region }

\subsubsection{Dependences of the He\,II line and nearby blends}
\label{heanal}

  In order to study the He\,II $\lambda$1640 line's behavior with 
respect to physical variables, we utilized the {\it SYNSPEC} line
synthesis code (Hubeny, Lanz, \& Jeffery 1994)
using LTE and non-LTE atmospheres for the effective 
temperature interval 21,000--29,000\,K at log\,g = 4 and for $\xi$= 5 
km\,s$^{-1}$. 
LTE models were taken from Kurucz (1993), while non-LTE models were taken 
from an extension of the OSTAR2002 grid for B stars (Lanz \& Hubeny 2006). 
Although the $\lambda$1640 line is an important transition complex in
the He$^{1+}$ atom, its proximity in wavelength 
to several iron-group lines of comparable strength is one
reason why it has been so little studied in B stars.  
The {\it SYNSPEC} program ameliorates this problem by permitting the
user to identify the primary lines by eliminating candidates from the input 
line library and seeing if a feature of interest has disappeared in the 
recomputed spectrum. The program provides the ability to convolve the
spectrum to mimic the effects of instrumental and rotational broadening.
We utilized these functions
to determine the contribution of the He\,II line to the 
aggregate ``$\lambda$1640 feature" as a function of stellar $T_{\rm eff}$ 
in spectra broadened by rotation.

\begin{figure}
   \centering
   \includegraphics[width=6.5cm,angle=90]{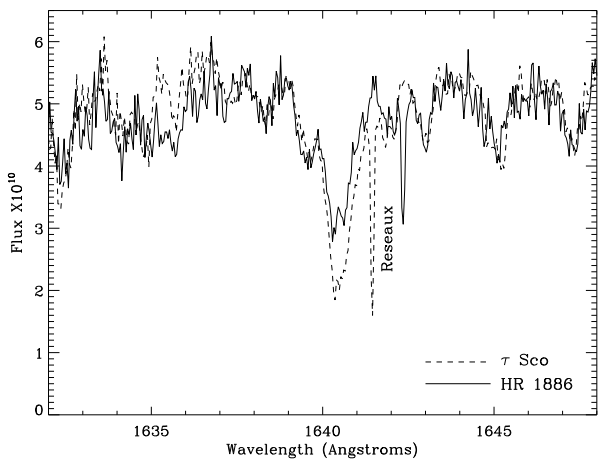}
%FIGURE 1 -
 \caption{ High dispersion {\it IUE} spectra of the $\lambda$1640
wavelength region of the sharp-lined stars $\tau$\,Sco (B0.2\,V) and
HR\,1886 (B1\,V).  The spectra partially resolve the He\,II 1640.4\,\AA~ 
line from the blend of Ni\,III/Fe\,II blend near 1640.0\,\AA. The wavelength
range depicted corresponds approximately to the velocity range exhibited in
most of the figures in this paper. The "Reseaux" title notes the presence
of a fiducial marker in the focal plane of the detector of the {\it SWP}
camera.  }
\label{atls}
 \end{figure}

  To show why a spectral line synthesis approach is important to the study
of $\lambda$1640, we exhibit in Figure\,\ref{atls} the wavelength region
surrounding this line taken from
{\it IUE} spectra of two B1 stars, $\tau$\,Sco and HR 1886. The spectra
of both stars are sharp-lined.  As expected, we see that the He\,II
line is stronger in $\tau$\,Sco ($T_{\rm eff}$ =30,200\,K, log\,g = 4.2; 
Hunter et al. 2005) than in HR\,1886 ($T_{\rm eff}$ = 23,300\,K, log\,g 
= 4.1; Lyubimkov et al. 2005). The figure also shows that the 
absorptions of nearby lines must be included in the measurement of the 
total strength of ``$\lambda$1640" in broad lined spectra. Our spectral
line syntheses have enabled us to identify these blends as follows.
Started from the blue edge of the aggregate, that the strong feature at
1639.4\AA\ is itself a blend of a Zn\,III ($\lambda$1639.42) and an Fe line.
The latter may be either Fe\,IV $\lambda$1640.40 or Fe\,II $\lambda$1640.40 
for spectra of type B0 or B2, respectively. Dominating the blue half of
the He\,II line is a feature at 1640.0\AA. For late O to B0.2-type stars,
this feature is a blend of excited Fe\,IV lines at $\lambda$1640.04
and $\lambda$1640.15.~ For B1--B2 stars the composition of this blend 
has shifted to Ni\,III $\lambda$1639.99 and secondarily to Fe\,II 
$\lambda$1640.15.~  Just to the red of the He\,II line, a strong
blend of Fe\,IV $\lambda$1640.78 line at type B0 gives way to a weak 
Fe\,II $\lambda$1640.86 line. 

 \begin{figure}
\centering
   \includegraphics[width=6.5cm,angle=90]{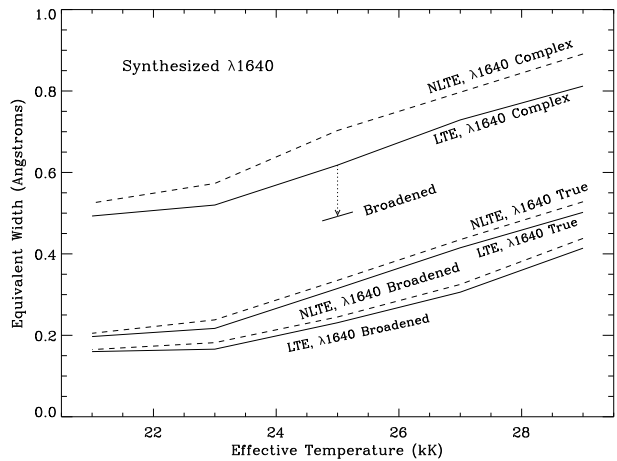}
%FIGURE 2 -
 \caption{
  The equivalent width contributions to the He\,II $\lambda$1640 blend as a
function of stellar effective temperature, computed from both LTE (solid line)
and non-LTE (dashed) model atmospheres and measured over the wavelength
interval $\lambda\lambda$1638.7--1641.5.  Pairs of solutions are shown for the
unrotated and broadened ($V_{rot}sin\,i$ =  300 km\,s$^{-1}$) profiles. 
The lower group of lines correspond to the 
equivalent widths found for the He\,II transitions alone.
Normal compositions and a microturbulent velocity of 5 km\,s$^{-1}$ are assumed.}
\label{he2comp}
 \end{figure}

  The fractional contribution of the He\,II line within the $\lambda$1640 
feature, along with the dependence of the strength of the total aggregate, 
is plotted in Figure\,\ref{he2comp}. This figure depicts the equivalent
width-temperature relations for three pairs of spectra computed with 
{\it SYNSPEC} for LTE and non-LTE atmospheres.  
In constructing the first (``True") pair of 
models, we have removed all lines from metal ions from the input line
library and computed the equivalent width between ``continuum" points 
at 1638.7\AA\ and 1641.5\AA.~ This is almost identical to the wavelength
interval selected by Peters (1990) for her measurements. Second, and beneath 
the ``True" relations in Fig.\,\ref{he2comp}, we show the relation for the 
same lines ``spun up" to a rotational velocity of 300 km\,s$^{-1}$ and 
measured according to the now lower ``continuum" points in the same
wavelength interval as in the first case.
A third pair of equivalent relations is measured with
the metal lines reinserted in the synthesized spectrum. The equivalent 
width for a fully broadened feature of the $\lambda$1640 aggregate is shown
for reference in the middle of the diagram for $T_{\rm eff}$ = 25,000\,K. 
Altogether, Fig.\,\ref{he2comp} brings out several characteristics of the
$\lambda$1640 feature. First, all relations run roughly parallel to one
another. Therefore, whichever relation one follows will undergo the same 
fractional change as one moves to a new effective temperature. For example,
it is a remarkable coincidence that the blends to the blue and red of He\,II 
maintain their strengths relative to He\,II through the early B spectral types.
Second, the loss of equivalent width from an undersetting of 
the continuum in a rotationally broadened spectrum is substantial, about
30\%. This affirms the necessity of measuring the strengths of the 
$\lambda$1640 aggregate in the same way for the range of stellar rotational 
velocities. Third, the contribution of the iron lines is almost half the 
strength of the total aggregate (for example, 0.30\AA~ of the total 
of 0.62\AA~ at $T_{\rm eff}$ = 25,000\,K).  
Fourth, the enhancement of the line strength due to non-LTE 
effects is very small.

% All told from 
% Fig.\,\ref{atls} and \ref{he2comp}, the difference in temperature between 
% $\tau$\,Sco and HR\,1886 leads to almost exactly the same percentage change 
%of equivalent width in the ``true" relation" as that measured (47\% vs. 45\%). 
% However, this
% agreement is almost certainly fortuitous. A comparison of the $\lambda$1640
% feature from $\tau$\,Sco is within measurement errors the same as for
% HR\,1886, even though the former is several thousand degrees hotter. Thus,
% it is in disagreement with our line synthesis results. Small changes 
% in log\,g could not affect the line strengths at this level, e.g., through 
% Stark broadening or the implied change in wind strength, and in any case 
% they would cause changes in the wrong sense. Thus, on the basis of
% these three well-studied sharp-line stars there appear to be other influences
%on the $\lambda$1640 line strength. (It is possible but not likely that errors 
% in determining the effective temperatures of these stars, which these authors
% derived primarily by Balmer line
% fits, are the sole cause.) We cannot identify the ``mystery parameter,"
% but the recent discovery of a complex magnetic field on 
% $\tau$\,Sco (Donati et al. 2006) raises the possibility that nonradial
% and time-varying wind configurations could be one factor.

\subsubsection{Statistical analysis of the $\lambda$1640 variations}
\label{statan}

   In order to undertake an quantitative analysis of the $\lambda$1640
variations, we first ``conditioned" the data, that is we standardized
the continuum levels and slopes (which changed over the {\it IUE} lifetime 
as the detector degraded) of the constituent spectra. We performed these 
steps by coadding all the spectra and choosing two generally line-free
regions across the order containing the He\,II line and resonance doublets
of interest. For this purpose we chose two velocity regions  (relative to
line center) at -1200\,--\,-700 km\,s$^{-1}$ and +350\,--\,900 km\,s$^{-1}$). 
The blue window of each spectrum was scaled relative to the mean.
The spectrum were then individually detrended relative to the mean again 
by an interactive computer routine
(generally by $\le$5--7\% from one end of our echelle order to the other). 
We then binned the spectra in wavelength by
a ratio of 2 to 1 pixels, thus making the value of each binned pixel
substantially independent of the values of its neighbors. Although in 
a few cases we have smoothed spectra for our plotting presentations, our
statistical tests described below were performed on the unsmoothed data.
To obtain an estimate of the r.m.s. noise level of our spectral comparisons
shown in Figures\,\ref{le9395}--\ref{19mions}, we used the median of the 
absolute value of the differences of spectra in the quasi-continuum windows.
The characteristic signal-to-noise ratios derived in this fashion were 
20--27 per binned pixel for pairs of spectra, $\sim$35 for comparisons 
of one spectrum against a seasonal average, and $\sim$70 for averages 
of two seasons represented by large numbers of spectra. 

  To determine the statistical significances of our trial $\lambda$1640
variations, we made the assumption that
the data noise is gaussian. Veteran users of {\it IUE} data will
recognize that this is formally a risky assumption for flux excursions of
perhaps 2 r.m.s. or more. However, since most flux differences within 
the profile do not exceed 1--1${\frac 12}$ r.m.s., the errors caused
by this assumption are unlikely to be serious.  We quantified the
statistical significances, ``$\sigma$," of the line strength variations 
with a computer a program we wrote that uses simple Monte Carlo approach. 
Our procedure 
was to define the wings of the entire profile and to use the r.m.s. level
determined above to determine the statistical likelihood of random
variations across the profile causing a difference in the absorption
anywhere in the profile by at least the observed amount. Notice that this
procedure estimates the significance level irrespective of ``sympathetic"
responses in `the C\,IV or Si\,IV lines. 
   
 As an initial check on our technique, we compared the fluctuations
of He\,II line profiles of the rapidly rotating B3\,V star
$\eta$\,UMa. The {\it IUE} satellite observed this star a total of 64
times with the SWP camera as a calibration standard 
during the interval 1978--1994. Using our Monte Carlo program, 
we searched for statistically significant line profile variations from the 
mean profile.  We found fluctuations neither over the whole profile 
or the red or blue halves greater than 1.6--1.7$\sigma$.  
Moreover, in the instances of greatest fluctuations from the mean there 
were no sympathetic responses in the C\,IV or Si\,IV lines.

  Because the ``eye" is a good judge of sustained flux variations
of several pixels, it is not surprising that we found
the overwhelming number of candidate variations we initiallly selected
turned out to be significant to at least the 3$\sigma$ (0.13\%) level. 
We relaxed this criterion only in Figure\,\ref{ec3}, in which evidence from a
simultaneous strong C\,IV variation in the same velocity range is overwhelming. 
Note also that our algorithm tests only changes in overall line strength. 
Thus, it is not an effective tool to measure the significance of high 
frequency variations of opposing signs across the profile.  
For this reason we withdrew two examples of possible ``emission
spikes" because they were not found to be 
statistically significant when tested against simulations over the whole 
line profile (typically ${\pm 300}$ km\,s$^{-1}$). In several cases the 
line strength contribution in one region of the profile overwhelmed a 
variation of opposite sign in another and nevertheless was 
significant over the whole profile. For example, for
Fig.\,\ref{2v8392} our annotated significances $\sigma$ refer to the 
{\it net} difference of the opposing contributions. In two  
of our examples  (Figs.\,\ref{mc1}, \ref{mc3})
opposite contributions of two segments of the line profile are about
equal. In these cases we will give the significances for the 
corresponding halves of the profiles.
Overall, it is likely that our procedures have excluded
several true variations of marginal significances.

\subsection{Selection of $\lambda$1640-Variable Be Stars}

  The impetus for this program 
was the example of $\lambda$1640 variations in the B2e star $\lambda$\,Eri 
and $\mu$ Cen. As discussed below, $\lambda$1640 and various optical He\,I 
lines in the spectra of these two stars undergo
rapid activity. This fact has led several authors to suggest
that magnetic fields play a role in this activity. Recently, Neiner
et al. (2003) have reported the detection of a rotationally modulated 
magnetic signature in $\omega$\,Ori. Thus, we will start our 
survey of He\,II line variability by discussing these three stars. 
Various authors
(e.g., ten Hulve 2004) have suggested that magnetic fields play a
role in aperiodic variability of the photospheric components
of the C\,IV and other resonance lines. We will therefore treat these
stars as well.

  To extend the search for He\,II line variability further, we 
surveyed all B0--B2.5 Be stars that the {\it
IUE} observed at high dispersion through the large aperture of the SWP 
camera at least 10 times.  This search netted a sample of 74 stars.
To this number we also added several early-Bn stars,
such as $\eta$\,UMa, that were observed many times, but none of them
exhibited He\,II line variations. 
Likewise, we note that some Bp stars exhibit variations in $\lambda$1640 
because of their heterogeneous He surface abundances (Bp stars such as
$\sigma$\,Ori\,E), and these are not included in our program. 
Our search is admitted not exhaustive, and it may
contain selection biases. However, note that rotational velocity was not 
a search criterion, except implicitly through our choice of Be stars.

\begin{table*}[ht!]
%\tablenum{1}
\begin{center}
\centerline{~}
\begin{tabular}{rrlrrrrr}  \hline

Name & HD Name & Sp. & Ref.  &  $Vsin\,i$ & Ref.  & $T_{\rm eff}$ &  Ref. \\
\hline
$\lambda$ Eri  & HD\,33328 & B2\,IVe & 1 & 325 & 1 & 24,000&2\\
$\psi$$^1$ Ori & HD\,35439 & B1\,Ve & 3 & 305 & 1 &    25,400 & 3 \\
$\omega$ Ori   & HD\,37490 & B2e\,III & 4  & 172 & 5 &  20,020 & 5 \\
19 Mon         & HD\,52918 & B1\,IV & 6   & 264 & 7 &    24,000 & 5 \\
HR 3186 & HD\,67536 & B2.5n(e) & 8 & 292 & 9 & 20,000 & 10 \\
$\mu$ Cen      & HD\,120324 & B2\,IVe  & 11 & 130 & 12    &  19,970 & 13 \\
$\eta$ Cen     & HD\,127972 & B1.5\,IVe & 14 & 350 & 4   &  21,860 & 15 \\
2 Vul          & HD\,180968 & B1\,IV & 6  & 332 & 16    &  25,000 & 10 \\
$\pi$ Aqr      & HD\,212571 & B1\,Ve & 6   & 250 & 17   &  25,000 & 17 \\
6 Cep          & HD\,203467 & B2.5e & 18  & 150 & 18 & 20,000 & 10 \\
\hline 
\end{tabular}
\caption{\label{}{Relevant parameters for program Be stars (ordered by right ascension)}}
\end{center}

{\bf Reference Key.} 
1: Abt 2002, 2: Hummel \& Vrancken 2000, 3: Lamer \& Waters
1987, 4: Slettebak 1982, 5: Neiner et al. 2003, 6: Lesh 1968, 7: Balona 2002,
8: Hanuschik et al. 1996, 9: Uesugi \& Fukuda 1970, 10: this paper, 11: Hiltner
et al. 1969, 12: Brown \& Verschueren 1997, 13: Rivinius et al. 2001,
14: Levenhagen et al. 2003, 15: Stefl et al. 1995, 16: Balona 1995, 17: 
Miroshnichenko et al. 2002, 18: Slettebak 1994.
\end{table*}

 Table\,1 lists 10 program stars for which we have found 
variable He\,II lines, representing a data sample
of 558 {\it IUE} SWP-camera echellograms.
The table also gives spectral types, $V\,sin\,i,$ and $T_{\rm eff}$ values
 according to the cited references. 
We have given preference to spectral types determined at high  
resolution and for velocities and temperatures to determinations by 
recent authors. Rough estimates of  $T_{\rm eff}$ 
for HD\,67536, 2\,Vul, and 6\,Cep are based on the observed He\,II line
strength, but these were not used in this work.

\section{Results}

\subsection{Magnetic candidates: $\lambda$\,Eri, $\omega$\,CMa, and
$\mu$\,Cen } 
\label{reslt}

\subsubsection{$\lambda$ Eridani } % --- \#9395, 2, 5,1  (drop \#3,4)  }

  $\lambda$\,Eri is a rapidly rotating B2\,IVe star that
has been studied extensively in the optical, UV, and X-range wavelength
ranges. Searches for velocity variations have produced no evidence that
the star is in a binary (Bolton 1982, Smith 1989). 
Since the discovery of H$\alpha$ emission in this star by Irvine (1975), 
this line has been observed to cycle between emission  and
absorption states.  Bolton (1982) first reported that the star
to be a periodic velocity variable.
These variations are now generally recognized to be due
to nonradial pulsations (NRP; e.g., Rivinius et al. 2003, but cf.
Balona \& James 2002). The star's broad spectral lines have prevented
the direct detection of a magnetic field. However, magnetic activity
might explain several types of activity. The first example of this,
``dimples," has been noted already. Second, Smith (2000) has noted 
the occasional presence of ``high velocity absorptions" in the He\,I  
$\lambda$6678 profiles. 
These events last several hours and have been interpreted as ejections 
of blobs, many of which return to the star (Smith, Peters, \& Grady 1991, 
Smith 2000; see also $\mu$\,Cen discussion). A third possible case for
magnetic activity is the observation of ``flows" across the
 He\,I $\lambda$6678 line profile over several hours (Smith 1989). This
suggests that the matter is channeled along prominence-like structures 
over the star's surface. A fourth example is the observation by the
{\it Rosat} satellite of a strong soft X-ray flare, lasting several hours
(Smith et al. 1993). Groote \& Schmitt 
(2004) have pointed to the similarity of this event and a flare observed
in the magnetic Bp star, $\sigma$\,Ori, E.

Evidence also exists for a periodicity or cyclicity of $\approx$475 days 
for the star's ``Be outbursts" (Mennickent, Sterken, \& Vogt 1998, Balona
\& James 2002).  Several observers (e.g., Peters
1990) have also reported increases in H$\alpha$ emission strength, and the 
appearance of  Discrete Absorption Components (DACs) of the C\,IV and
Si\,IV resonance doublets at about -900 km\,s$^{-1}$.  
%It remains to be
%added that during one disk ejection episode infall events have been
%observed as redshifted emissions from a circumstellar ring or disk around 
%this star (Smith, Peters, \& Grady 1991).

 \begin{figure}
\centering
   \includegraphics[width=6.5cm,angle=90]{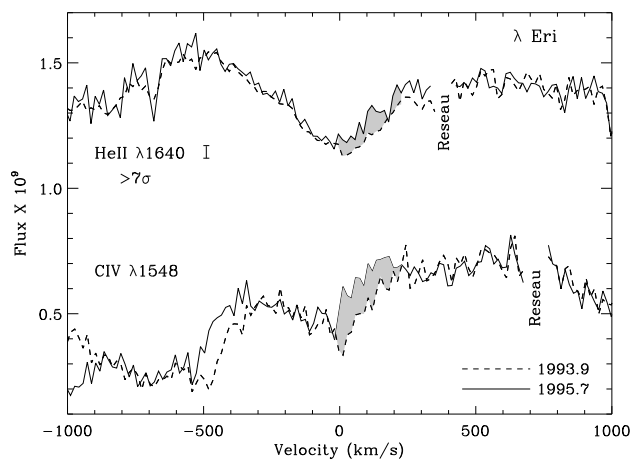}
%FIGURE 3 -
 \caption{{\it IUE} spectra of $\lambda$\,Eri obtained from observations
during the epochs 1993.7--1994.1 and 1995.8 for the He\,II and Si\,IV
doublet.  The 1993-4 spectra are smoothed
over 2 points in this plot.  The shaded areas indicate the variations of
the photospheric components; significance on He\,II is for this shaded area.  
In all plots, we assume a centroid wavelength
1640.3 for the $\lambda$1640 aggregate feature.  In this convention the
He\,II feature centroid will correspond to a velocity of about
+55 km\,s$^{-1}$.  Unless otherwised noted, the velocity system for the
C\,IV doublet is referenced to the $\lambda$1550.8
component. 
}
\label{le9395}
 \end{figure}

  We found variations over both short and long timescales in our 
examination of the He\,II $\lambda$1640 line of $\lambda$\,Eri in 
the {\it IUE} archives. As a start, we note
that the He\,II line equivalent widths 
decreased substantially, and with almost no overlap between their
respective ranges between the epochs 1993.79--1994.06 
and 1995.7. A comparison of the mean profiles during these times
is depicted in Figure\,\ref{le9395}.
This plot shows that the He\,II line is either partially filled in on the
red side (0 to +300 km\,s$^{-1}$), or Doppler shifted to to the blue.  
This interpretational ambiguity is decisively settled by the filling in 
of the red wing of the C\,IV doublet at 0--240 km\,s$^{-1}$ in this figure. 
A similar difference is found in the Si\,IV doublet (not shown).
 \begin{figure}
\centering
   \includegraphics[width=6.5cm,angle=90]{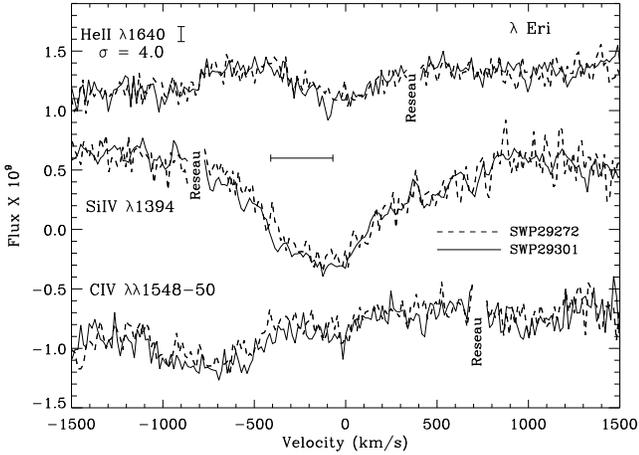}
%FIGURE 4 -
 \caption{Comparison of {\it IUE} observations SWP 29272 and 29301 (2.0
days later) on $\lambda$\,Eri for He\,II, Si\,IV $\lambda$1394 line and
the C\,IV doublet in the spectra.  The comb indicates the velocity range
over which there are correlated variations.
}
\label{le2}
 \end{figure}

 \begin{figure}
\centering
   \includegraphics[width=6.5cm,angle=90]{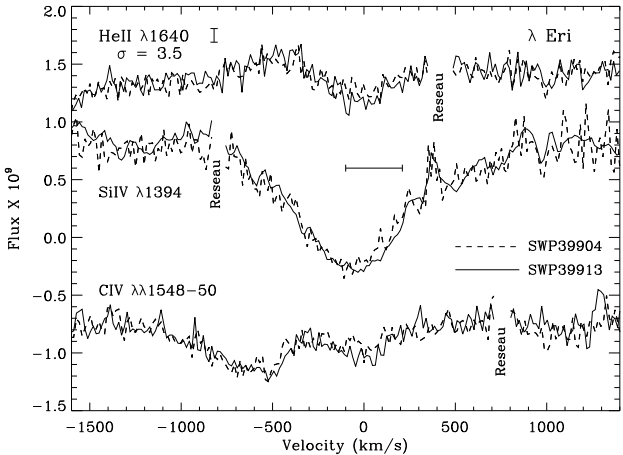}
%FIGURE 5 -
 \caption{Comparison of {\it IUE} observations SWP 39904 and 39911 (19
hours later) for He\,II, the C\,IV doublet and the Si\,IV $\lambda$1394
line in $\lambda$\,Eri. This figure shows typical rapid weakenings of
the He\,II line in this star.
}
\label{le5}
\end{figure}

\begin{figure}
\centering
   \includegraphics[width=6.5cm,angle=90]{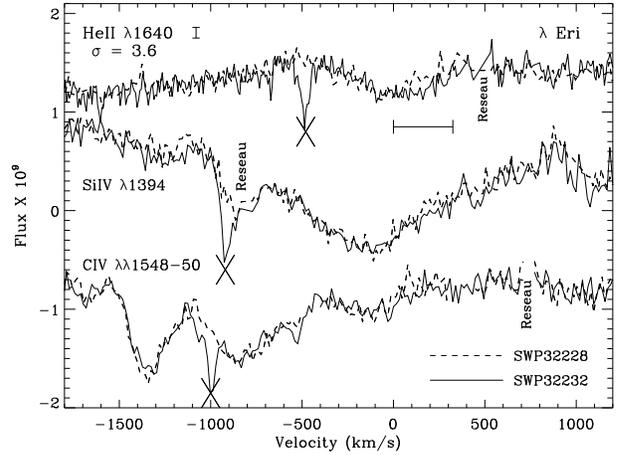}
%FIGURE 6 -
 \caption{Comparison of {\it IUE} observations SWP 32228 and 32232 (3.9
hours later) for the He\,II, the C\,IV, and Si\,IV $\lambda$1394
lines in $\lambda$\,Eri. This figure shows typical rapid filling in of
the red wing of the He\,II line in this star. The X-marked features in
all three features are instrumental (``missing minor frames").
% resulting from a brief telemetry interruption during read out.
}
\label{le1}
\end{figure}

  The {\it MAST/IUE} archive contains 145 SWP high-dispersion observations of
$\lambda$\,Eri distributed over a large range of timescales, including 
monitoring campaigns in 1982 and 1996. We have found rapid variations 
of the He,\,II, Si\,IV and C\,IV lines during these times. Figures \ref{le2} 
and \ref{le5} document changes over the central
region of the photospheric profile. The intervals between 
these two pairs of observations are 48 and 19 hours, respectively.
Fig.\,\ref{le5} is especially interesting because it compares the 
profiles during a pair of dimple-active and -inactive states (Smith et 
al. 1996; see Figs.\,5 and 6). As opposed to the He\,I lines, the 
He\,II respond to dimples by small {\it weakenings.} 
In Figure\,\ref{le1} we show variations over 4-hour in the red 
wings of the He\,II, C\,IV and Si\,IV $\lambda$1394 lines. Because the
increased flux may not exceed the continuum level, we do not know if this
is due to emission or to a weakening of absorption.

\vspace*{-.15in}

\subsubsection{$\omega$ Orionis } %  -- \#5, 1, 2 }  

   $\omega$ Ori is a typical B2e star that seems to oscillate between between 
B-normal and Be H$\alpha$ emission states. So far, these oscillations seem
to be cyclical rather than strictly periodic. The high-velocity (wind)
components of the resonance lines often exhibit large changes. The profiles
are relatively narrow for a classical Be star. This fact has permitted the 
discovery of a weak dipolar surface magnetic field  that modulates on a
rotational period of 1.29 days (Neiner et al. 2003).  Given the expected
radius of a B2 main sequence star, this period implies that 
we view this star from an intermediate aspect.
Neiner et al. (2003) also reported an enhancement of nitrogen from an 
optical line.  C\,IV variations inform us that the wind of 
this star can be variable on timescales as short as 1${\frac 12}$ hours. 
Such variations suggest that localized and hence anisotropic changes in the 
wind in the rapidly accelerating zone occur close to the star (Sonneborn 
et al. 1988).
  
   The {\it IUE} archive includes 189 SWP high-resolution observations of
$\omega$\,Ori. Of these, 110 are included during intensive campaigns in 1982 
and 1996 and intermittent monitoring in 1983.  According to ground-based 
polarization studies, 
the star underwent a strong, rapidly evolving outburst in 1983 
(Sonneborn et al. 1983). This event was accompanied by the emergence of even 
stronger DACs in the C\,IV lines, though these were shifted to lower velocities. 
The star's H$\alpha$ line was in strong emission in December, 1996, and
by 1999 its equal $V$, $R$ emissions were still unchanged (Peters 2005). 
Thus, it is likely that when the 1996 campaign was conducted, 
$\omega$\,Ori was likewise in a Be active state.
Although the star's H$\alpha$ line showed strong emission 
during the 1982--3 episode, there are apparently no published accounts of
its status during early 1996. However, Neiner et al. (2003) documented that
the emission was moderately strong in 1998 and declining through 1999. 
For completeness, we note that the 1996 spectra of $\omega$\,Ori
exhibit stronger N\,V doublet absorptions than during 1982. This is among 
the few cases in our study for which we could find a correlation 
between these He\,II and N\,V temperature indicators. 
The difficulty of seeing the correlation for other stars 
is mainly due to the weakness of N\,V lines in their spectra.

 \begin{figure}
\centering
   \includegraphics[width=6.5cm,angle=90]{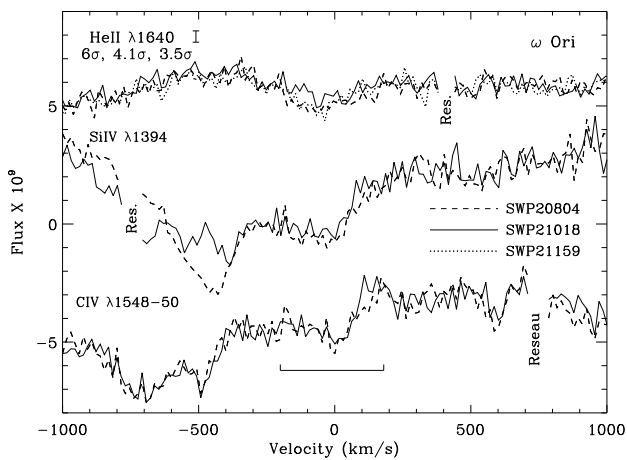}
%FIGURE 7 -
 \caption{{\it IUE} observations of $\omega$\,Ori during 1983 that
show variations of the He\,II, Si\,IV and C\,IV lines.
The interval between the solid and dashed line observations is 16 days.
The dotted line (He\,II only) shows the observation SWP\,21159 13 days
after the second observation. The significances quoted is for differences
between the first and second, second and third, and first and third
He\,II observations, respectively.
}
\label{oo83}
\end{figure}

  We found several examples of rapid and long-term variability in 
the He\,II line variability for this star.
Starting again with the long-variations, we noticed systematic differences 
in the mean photospheric profile strengths for 1982 and 1996 observations. 
The 1982 profiles exhibit a filling in of the red wing
and a tapered blue absorption wing.
The variations in the He\,II line of $\omega$ Ori are not always replicated 
in the C\,IV and Si\,IV  doublet lines, even though the doublets
show substantial {\it inter alia} variations at high velocities.

Sonneborn et al. (1988) discussed the wind activity in this star 
during 1982--3. Figure \ref{oo83} exhibits one observation, SWP\,21018, 
discussed in their paper and another obtained 16 days earlier. During this 
time the He\,II profile showed substantial activity over sometimes broad, 
and at other times narrow, wavelength ranges. 
The Si\,IV and C\,IV doublets exhibited
apparent incipient emission fluctuations in their red wings during these
times. In the blue/central parts of the line, the Si\,IV doublet changed
very little and the C\,IV lines not at all.  Some 13 days later, the He\,II 
line shows the same filling in on the red side as for SWP\,21018, but 
the blue side of the line again shows full-strength absorption.  
An important interpretation from these
comparisons of He\,II and resonance line 
variations is that changes observed at the base of the wind do not 
necessarily correlate well with those further out in the flow.

 \begin{figure}
\centering
 \includegraphics[width=6.5cm,angle=90]{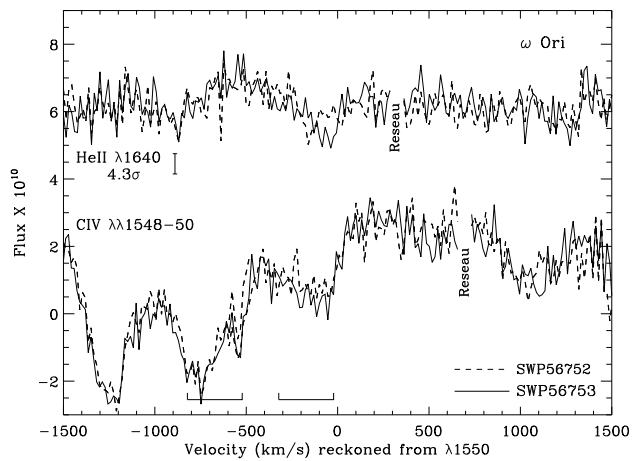}
%FIGURE 8 -
 \caption{{\it IUE} observations of $\omega$\,Ori over a 12 hour
interval in 1983, exhibiting variations of the He\,II, Si\,IV and
C\,IV lines. Errors on He\,II are taken in the interval 0--350 km\,s.
}
\label{oo5}
\end{figure}

  These lines also exhibit variations on a rapid timescale of 12 hours; see 
Figure\,\ref{oo5}. In this case the central core of $\lambda$1640 (solid
line) has deepened while the red wing has filled in. The
C\,IV doublet shows an overall weakening over the whole photospheric
components. The Si\,IV lines exhibit no change.  We speculate that
because $\omega$\,Ori has a magnetic field,
these rapid variations might arise from dissipative magnetic processes 
in the outer atmosphere.

% \begin{figure}
%\centering
% \includegraphics[width=6.5cm,angle=90]{f9.eps}
%GIGURE 9 -
% \caption{A comparison of the He\,II and C\,IV lines over an
% interval of only 2.8 hours during 1983 for $\omega$\,Ori.
% }
% \label{oo2}
% \end{figure}

\vspace*{-.15in}

\subsection{Stars with possible $\lambda$1640 red wing emission}
\label{emis}

\subsubsection{$\mu$\,Centauri  }   % 1,3,2}   

  This B2e star has been extensively observed and shows a rich activity
in its light curve, H$\alpha$, and other spectral lines. The light
curve undergoes sporadic brightenings of up a few tenths
of a magnitude (Baade et al. 2001) for reasons unknown.
The H$\alpha$ emission component exhibits activity episodes over 
a variety of amplitudes and timescales (e.g., Hanuschik et al. 1993). 
% From a near-IR spectroscopic survey, Baade (1992) concluded that there is
% no evidence for radial velocity variations or for a cool companion. 
The star's spectral lines are sharp, suggesting that it is viewed at a low 
inclination. $\mu$\,Cen has been extensively monitored spectroscopically.
Rivinius et al. (1998b) reported that its line profile variability can
be decomposed into six nonradial pulsation periods. These modes cluster at 
0.51 and 0.28 days. Rivinius et al. (1998a) have attempted to interpret
the Be outburst in this star as an outcome of nonlinearities associated
with the beating of these modes. In addition to activity associated with
intermodal beating, aperiodic activity also appears to be present. Peters 
(1984b) noted that an optical He\,I line exhibited a
rapidly appearing absorptions at velocities outside the photospheric 
profile over several minutes. Such transients, since dubbed
``high velocity absorptions" seem to represent 
discrete ejections of blobs (Rivinius et al. 1998). 
%The ejected blob spiral 
%outward at first and then probably disperse and contribute to building an
%axisymmetric Be disk (Smith 2000, Rivinius 2005). 
Peters (1998) 
has also documented evidence of rapid, large-amplitude emission variations
along the $\lambda$6678 profile over several hours, suggesting that matter
is channeled in arc-line prominences.

 \begin{figure}
\centering
 \includegraphics[width=6.5cm,angle=90]{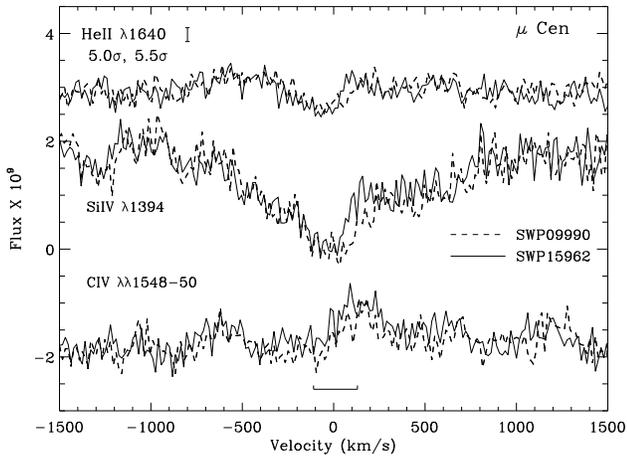}
%FIGURE 9 - (old figure 10)
 \caption{A comparison of $\mu$\,Cen's He\,II and C\,IV lines in
1980 and 1982 for $\mu$\,Cen. Note the weak red wing emission.
Errors on He\,II refer to blue and red halves of the profile, respectively.
}
\label{mc1}
\end{figure}

 \begin{figure}
\centering
 \includegraphics[width=6.5cm,angle=90]{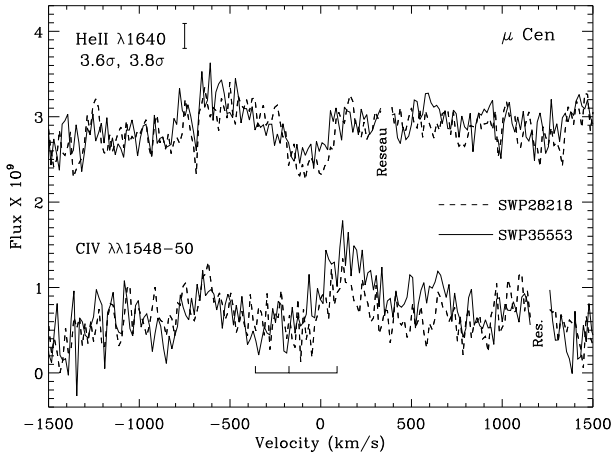}
%FIGURE 10 - (old figure 11)
 \caption{A  comparison of the He\,II and the C\,IV lines for the interval
April 1986 -- February 1989 for $\mu$\,Cen.  
%The 1989 observation has been smoothed over 2 points. 
Note the weakening of the He\,II line coincident
with the formation of red wing emission in C\,IV.
Errors on He\,II refer to blue/central and red portions of profile (see comb).
}
\label{mc3}
\end{figure}

   In our survey of the 34 available  SWP echellograms for
$\mu$\,Cen, we found several variations of the $\lambda$1640 line. 
This included a narrowing of the  He\,II, Si\,III, Si\,IV, and Al\,III lines
from September 1980 to February 1981 reported by Peters (1984a).  However,
whereas Peters interpreted these differences in terms of line broadening
during an H$\alpha$ emission episode, we believe these differences should
be interpreted as a weakening of absorption and emergence of P Cygni emission
component during the {\it later epoch,} making the profiles appear narrow. 
These differences can be seen in Figure\,\ref{mc1}. Any possible doubt that
the faint bump in the red wing of the C\,IV doublet are emission in this 
figure is dispelled by its behavior in Figure\,\ref{mc3}. 
This plot compares observations of the He\,II and C\,IV lines made in 1989 
and 1996. The 1989 profiles are weak and narrow. 
Although there is much activity in the red wings, the photospheric 
components of C\,IV show no variations.
Just as for $\lambda$\,Eri and $\omega$\,Ori, we were surprised to find
that relatively large variations in the He\,II line of $\mu$\,Cen can occur 
on short as well as long timescales. Although we have found convincing 
evidence for at least four cases in the He\,II line, residual red wing emission 
is actually more common for the C\,IV and 
Si\,IV doublets (Fig. \ref{mc1}).  Moreover, short-term variations can occur
in resonance lines of less excited ions like Si\,III and Al\,III 
(Peters 1984a). This suggests that the region of the wind affected extends 
further downstream than is the typical for other stars in our sample.

% \begin{figure}
% \centering
%  \includegraphics[width=6.5cm,angle=90]{f12.eps}
%GIGURE 12 -
%  \caption{A comparison of the He\,II and C\,IV lines for $\mu$\,Cen
% over a 3.7 hour interval on 1985 June 28.
% }
% \label{mc2}
% \end{figure}

\vspace*{-.15in}

\subsubsection{6 Cephei } %  -- \#}

  Like $\mu$\,Cen, 6\,Cep is a B2.5e star with unusually narrow spectral
lines for a Be star. Pavlovski et al. (1997) monitored the star's optical 
flux but were unable to find optical continuum variations. 
In contrast, both its optical and UV spectrum are variable. 
The C\,IV resonance lines exhibit strong variations
during oscillations of its wind state (e.g., Barker and Marlborough  
1985, Grady, Bjorkman, \& Snow 1987; ``GBS").  Abraham et al. (1993) have 
speculated that the star's wind is responsible for the 
creation of a ``stellar wind bubble," which they were able to image with
the {\it IRAS} satellite. Koubsky et al. (2005) have found a period of 1.621 
days in the optical line profiles of 6\,Cep. 
These authors believed that they could 
be attributed either to nonradial pulsations or a corotating disturbance or
cloud over the star. However, we favor the pulsation alternative because 
in our present study we can see  variations in lines arising from moderately
excited exitation states, as would be expected in the photosphere.

\begin{figure}
\centering
 \includegraphics[width=6.5cm,angle=90]{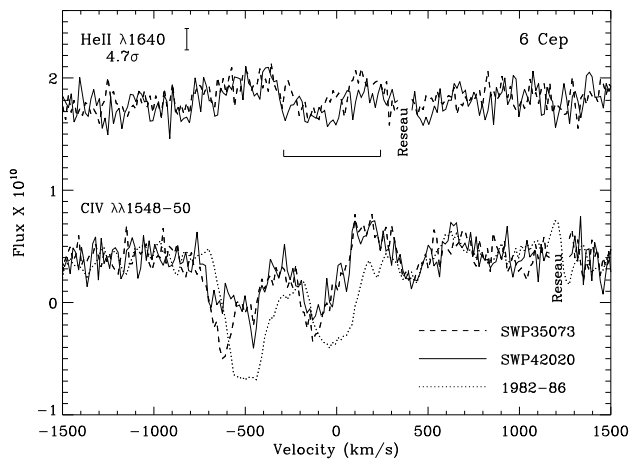}
%FIGURE 13 - (old figure 11)
\caption{A comparison 6\,Cep's He\,II and C\,IV lines for epochs 1989.0
and 1990.5. Note the inactivity of the red wing
emission in the C\,IV line. The wind was in a different state during 
1982-86 (C\,IV, dotted line).  }
\label{6c1}
\end{figure}

   The {\it IUE} archive contains 34 SWP echellograms of 6\,Cep. 
From our inspection, we suspect that a few profiles undergo small amplitude
variations. However, our statistical tests of these disclosed that only one 
was significant. A comparison of He\,II and the C\,IV doublet
is shown in Figure\,\ref{6c1} for a pair of observations taken in 1989
and 1990. In this example, the He\,II line shows a general weakening
of the photospheric profile and a distinctly raised red wing.
The corresponding profile of the C\,IV doublet is consistent with true 
emission:  its red wing is raised above 
the continuum level. Although the red wings of the C\,IV exhibit no changes 
during this time, the profiles develop a narrow absorption at about 
-150 km\,s$^{-1}$. This is just one of a few examples shown in the present
paper of different types of simultaneous variations of He\,II and C\,IV. 
The example in Fig.\,\ref{6c1} also
adheres to our more general finding that a filling in 
of red wing emission in the He\,II line tends to accompany strong
absorption at low negative velocity in the C\,IV doublet. To give some
perspective as to how these C\,IV profiles differ from the ``norm," we
exhibit in Fig.\ref{6c1} the profiles of both pairs of observations from
1989 and 1990 and the mean profile for 1982-6 (dotted line). 

\vspace*{-.15in}

\subsubsection{HD\,67536 } %  -- (c4hemontagev) }        

  HD\,67536 is a somewhat understudied variable B2.5-B3n star. On some 
occasions its Balmer lines show emission (Hanuschik 1996). 
The structure and significant variability of its C\,IV lines is similar 
to that found in 6\,Cep (GBS). Using this star as a prototype, ten Hulve 
(2004) and Henrichs et al. (2005) have identified a class of ``magnetic"
candidates based on the presence of variability between the rotational 
velocity limits ${\pm}V\,sin\,i$ in the line profiles of this doublet. 
This working definition is reminiscent of the
characteristic periodic absorption/emission behavior of these lines noted
by Shore and colleagues (e.g., Shore \& Brown 1990). However, the datasets
for most or all of the 24 Be stars so characterized are too sparse
to determine whether these variations are periodic. 

\begin{figure}
\centering
 \includegraphics[width=6.5cm,angle=90]{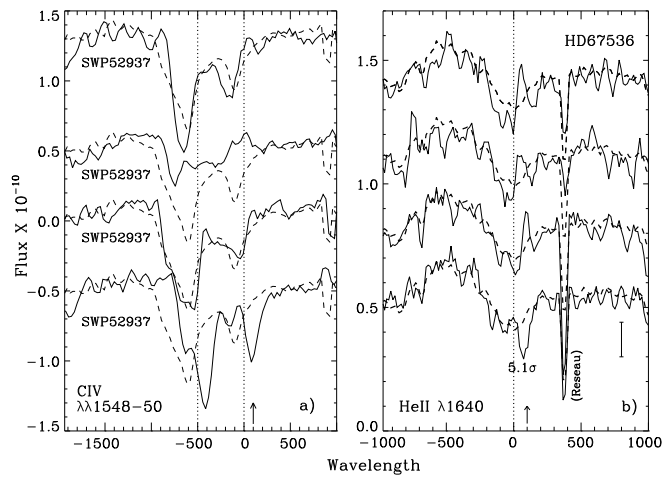}
%FIGURE 12 - (old figure 14)
\caption{A montage of four variations of the C\,IV and He\,II lines in
1983, 1986, 1994.7, and 1994.9 for HD\,67536.  The dashed line represents
the mean of the available 22 spectra. The photospheric components of the
C\,IV doublet of observation SWP\,27503 (bold line) are nearly
completely filled in.  The average spectra (dashed plot) are binned to 4
instead of 2 pixels. Note the slightly sharp, red displaced features (in
the first three cases in emission) in these lines. ``5.1$\sigma$" (panel b, 
fourth spectrum) refers only to the variation of the sharp absorption.
}
\label{hd675v}
\end{figure}

  In examining the 22 available {\it IUE} SWP echellograms for 
HD\,67536, we found that variations at velocities above -600  km\,s$^{-1}$
are readily apparent in the C\,IV and Si\,IV lines.
Figure\,\ref{hd675v} exhibits four spectra obtained at epochs 1983, 1986, 
1994.7, and 1994.9 along with the mean of all observations.
Interestingly, the He\,II profiles in the first three exposures have a
P\,Cygni-like character, which includes a narrow emission spike at about 
+100 km\,s$^{-1}$. (Because the profile variations are so complicated we
have not attempted to determine their statistical significances.)
The C\,IV lines also show arguably weak emission at 
this velocity. Indeed in the SWP\,27503 observation the doublet components 
are almost completely filled in out to the DAC at -300 km\,s$^{-1}$.
Incidentally, this peculiar phenomenon cannot be explained by
small changes in the $T_{\rm eff}$ over the whole stellar surface because
the iron-group lines seem unaffected.
Judging from previous exposures, the C\,IV line spectrum was in this 
filled-in state for at least 4 days when SWP\,27503 was recorded.
In the fourth example, depicted in Fig.\,\ref{hd675v}, an observation taken
in 1995.1, a narrow absorption appears, again at +100 km\,s$^{-1}$, in
C\,IV and He\,II lines. We not know if it is significant that each of 
these events occurs at nearly the same velocity.

\subsection{Stars with $\lambda$1640 blue-wing absorption }
\label{baswnd}

  In contrast to the previous Be stars, spectra in the second half of
our sample exhibit variations that are usually correlated with
blue wing variations in C\,IV and/or Si\,IV, and occasionally even Si\,III
resonance lines. 

\vspace*{-.15in}

\subsubsection{$\psi$$^1$ Orionis } %  -- \#2, 1 } 

$\psi$$^1$ Ori is in many ways a typical early-type Be star.
The spectral temperature diagnostics, 
including $\lambda$1640 and resonance line wind features, are
consistent with this classification. Like many other Be stars, it 
exhibits cyclic H$\alpha$ emission episodes. This emission was strong 
during the late 1970's and early 1980's (Barker 1983) when 
Lamers \& Waters (1987) estimated the mass loss of $\psi$$^1$ Ori 
to lie in the range 10$^{-8}$--4$\times$10$^{-7}$ M$_{\odot}$ 
yr$^{-1}$ from {\it Copernicus,} {\it IUE,} and {\it IRAS} data. This
is among the highest mass loss rate range noted in these authors' sample. 

  The {\it IUE} archive contain 26 SWP camera observations of this star.
These are well enough distributed to give a sense of both rapid and long-term 
variations. From these data we found no evidence of rapid variability.  
Even over the long term, the fluctuations of this line are small compared 
to the moderate amplitudes of C\,IV activity. 
These variations extend over all possible velocities. Curiously, 
a strong ``DAC-like" feature is present in the C\,IV complex at about 
-200 km\,s$^{-1}$ in all the observations. 

\begin{figure}
\centering
 \includegraphics[width=6.5cm,angle=90]{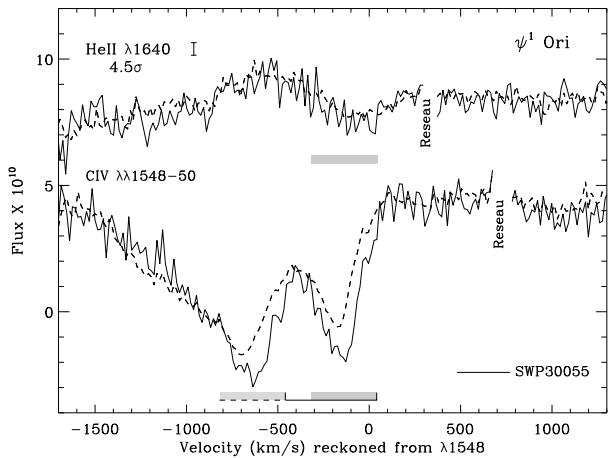}
%FIGURE 13 - (old figure 15)
 \caption{The comparison of the stronger than unusual
He\,II and C\,IV spectra for $\psi^1$ Ori obtained at 1987.2
relative to mean profiles.
  }
\label{po2}
\end{figure}

\begin{figure}
\centering
 \includegraphics[width=6.5cm,angle=90]{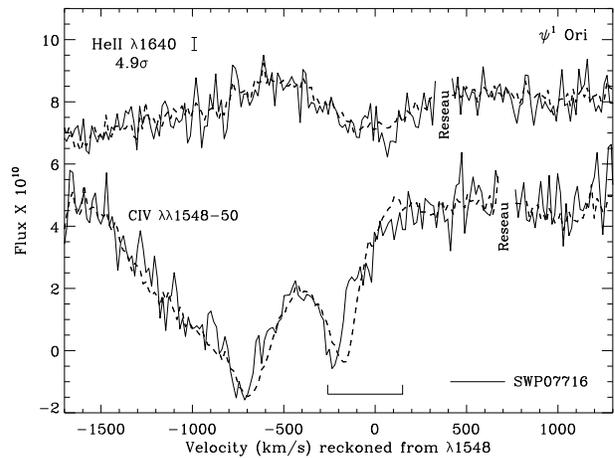}
%FIGURE 14 - (old figure 16)
 \caption{A comparison of the He\,II and C\,IV spectra
for $\psi^1$\,Ori at the epoch 1980.0 relative to the mean profiles. The
single spectrum shows a weak sharp absorption feature at 100 km\,s$^{-1}$.}
\label{po1}
\end{figure}

  The He\,II variations in $\psi$$^1$ Ori seem to bridge the red-central 
profile activity seen in the examples discussed above and for the
remaining stars in our sample. Up until the spectra of this star, 
we have not encountered He\,II variations in the blue wing. In fact, 
our first example, depicted in Figure\,\ref{po2}, exhibits 
an increased absorption over the range +50 to -300 
km\,s$^{-1}$.  The C\,IV lines show a similar variation in their blue 
wings, but they also exhibits a weakening at high velocities 
of -1000 to -1500 km\,s$^{-1}$. 
Figure\,\ref{po1} exhibits a second example of  He\,II
variability. On this occasion absorptions are present, a narrow feature
at +100 km\,s$^{-1}$ and a broad one in the blue wing. The narrow but 
not the broad feature is present in the corresponding C\,IV observation. 

\vspace*{-.15in}

\subsubsection{$\eta$ Centauri } %  \#3,2,1 }

  A rapidly rotating B2e star, $\eta$\,Cen has been the subject of much recent 
study. Its optical and UV lines and UV continuum
undergo regular short-term variations with
a dominant period near 0.64 days (Leister et al. 1995, Stefl et al. 1995,
Peters \& Gies 2000).  In addition, the star's H$\alpha$ line exhibits
strong oscillations over long timescales (Dachs et al. 1986).  Rivinius (2005)
has suggested that the star ejects blobs that develop into ring-like structures.

   Inspection of the C\,IV lines discloses considerable ``slow"
variability in the range -1500  to $\sim$+200  km\,s$^{-1}$.
The line cores show a peculiar double-lobed
structure, and the core centroid positions vary between -50 and
-250  km\,s$^{-1}$. Although the red wings of the C\,IV doublet are typically 
filled in by emission, the changes in their profiles are small. To a
lesser extent, these statements also apply to the Si\,IV doublet and
even the Si\,III $\lambda$1206 line.

%\begin{figure}
%\centering
% \includegraphics[width=6.5cm,angle=90]{f15.eps}
%%FIGURE 15 - (old figure 17)
% \caption{A comparison of the overall mean He\,II and C\,IV spectra 
%with the average of a spectrum of $\eta$\,Cen obtained in 1991.2
%}
%\label{ec2}
%\end{figure}

\begin{figure}
 \centering
  \includegraphics[width=6.5cm,angle=90]{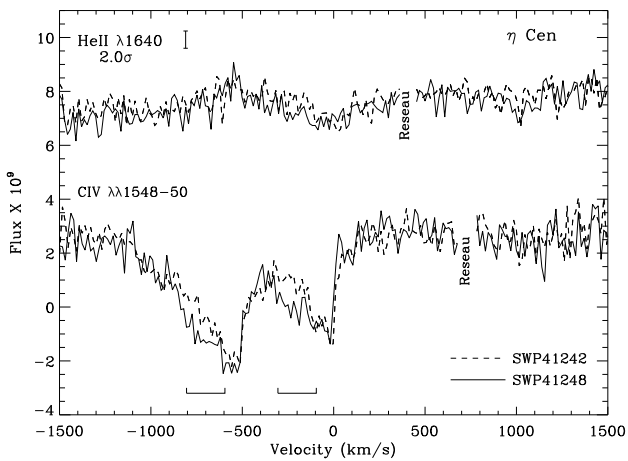}
%FIGURE 16 -
  \caption{A comparison of the He\,II and C\,IV spectra during the
 1991.2 campaign on $\eta$\,Cen.  Notice the variation of the blue wings over
 over a 7.4 hour interval. 
 Our significance acceptance criterion was relaxed
 in this one case because of the correlation with the strong C\,IV variation.
  }
 \label{ec3}
 \end{figure}

\begin{figure}
 \centering
 \includegraphics[width=6.5cm,angle=90]{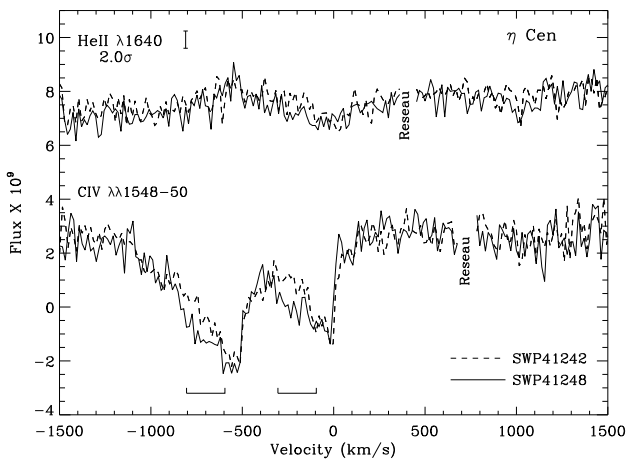}
%FIGURE 17 - (old figure 19)
 \caption{A comparison of the He\,II and C\,IV lines during the
1991.2 campaign on $\eta$\,Cen. Note the variation of the red wings
over a 3.9 hour interval.
  }
 \label{ec4}
 \end{figure}

  The {\it IUE} obtained 28 {\it SWP} echellograms of this star 
from 1983 through 1991. 
These were roughly evenly distributed between 
two observing campaigns during each epoch.
The He\,II line shows correlated variations with
C\,IV in several instances. First, we noticed long-term differences in
the sense that the wings of both lines were more depressed in 1991.
Rapid variability is also evident, and we display two examples. 
Figure\,\ref{ec3} exhibits variations in these lines over the same velocity
range and over an interval of 7.4 hours. During this time the He\,II
line and the C\,IV doublet developed a low-velocity absorption at
about -200 km\,s$^{-1}$.
Figure\,\ref{ec4} shows a weakening of emission over an
interval of 7 hours. Taken together, these two examples are the
only cases we have of activity in the blue and red wing of $\lambda$1640
line over a time interval that is not much longer than the star's
rotational period (1.5 days).

\vspace*{-.15in}

\subsubsection{$\pi$ Aquarii } %   1,2 }

   $\pi$\,Aqr is an active B1e star and is also a double-lined, 
84-day spectroscopic binary. The component mass ratio is 0.16, which implies
that the secondary is an early-type A star (Bjorkman et al.  2002). 
Detailed fits of the H$\alpha$ emission profiles suggest that we view this
star at a high inclination, i.e., $i$ $\simeq$ 70$^o$ (Hanuschik et al. 
1996). The X-ray flux of this star is high for a Be star, indeed about 
one half that of the X-ray anomalous Be star $\gamma$\,Cas. Also, unusually
for a Be star, high energy emission has been detected by the {\it EUVE} 
satellite (Christian et al. 1999). Moreover, its 
IR flux is quite variable, even for a Be star. 
% These facts have led several 
% authors to regard $\pi$\,Aqr as an interacting binary. 
Bjorkman et al. have suggested that the star's variable H$\alpha$ emission
can be used to measure a time-dependent mass transfer from the secondary star 
on to the Be star's disk. (This assumes that the disk is formed by binary
accretion and not by decretion, as is ordinary for Be stars.)
The abnormal absorption strength of the He\,I $\lambda$5876 is consistent
with the formation of part of this line close to the Be star's surface,
perhaps in the inner region of the disk.

   The {\it IUE} observed $\pi$\,Aqr 23 times during 1978--9 
and 1985--91 with the SWP camera. Unusually for a Be star, 
the N\,V doublet is not only present but strikingly variable. 
Ringuelet, Fontenla, \& Rovira (1981) have reported emission in 
these lines, and we attribute this to mass transfer from the companion
to the Be star. The
resonance lines of C\,IV, Si\,IV, and Si\,III exhibit a characteristic
range and type of variability for an active Be star. Moreover,
because its measured mass loss rate of $\approx$2.5$\times$10$^{-9}$  
M$_{\odot}$ yr$^{-1}$ (Freitas Pacheco 1982, Snow 1981) is typical for 
a Be star, we believe that the variable blue wings of these lines
are due to fluctuations in the Be star's wind.
It remains to be added that the Be star's optical lines reveal the 
presence of traveling bumps due to a 1.88 hour oscillation (Peters \&
Gies 2005; the grayscale in this paper offers an unusually clear 
depiction of the increased acceleration of NRP bumps at the edges 
of the line profile).

\begin{figure}
 \centering

 \includegraphics[width=6.5cm,angle=90]{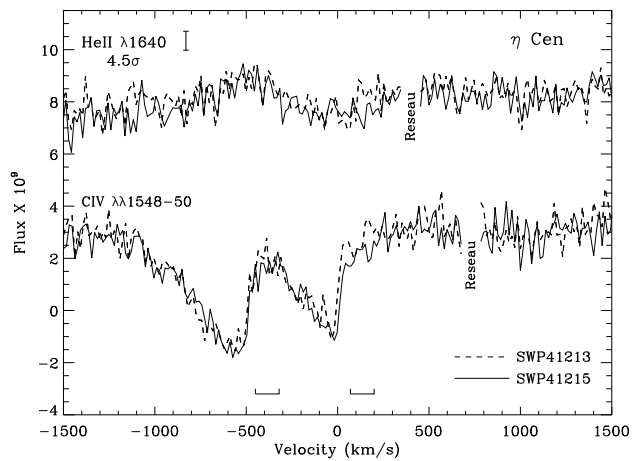}
%FIGURE 18 - (old figure 20) 
 \caption{A comparison of the He\,II, C\,IV, Si\,IV, and Si\,III spectra
 between epochs 1979.5 and 1979.2 for $\pi$\,Aqr.
   }
 \label{pa1}
 \end{figure}

\begin{figure}
 \centering
 \includegraphics[width=6.5cm,angle=90]{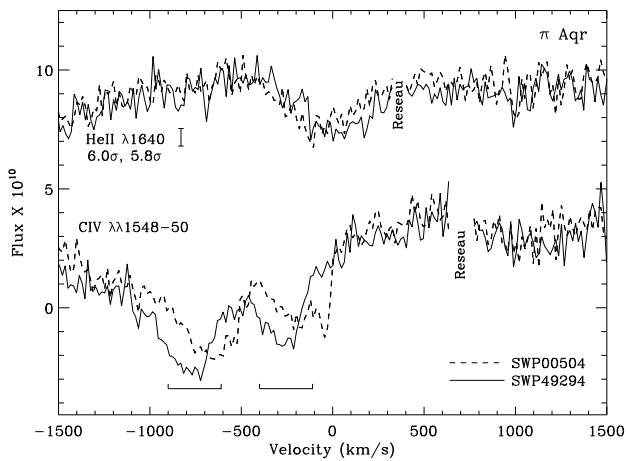}
%FIGURE 19 - (old figure 21)
 \caption{A comparison of the He\,II and C\,IV spectra between
 epochs 1988.4 and 1993.9 for $\pi$\,Aqr.
Errors on He\,II are given for the blue wing (comb interval) and 
red wing, respectively.
  }
 \label{pa2}
 \end{figure}

  This star's He\,II line exhibits remarkable blue-wing
strengthenings which track strengthenings of the
strong blue wings of the C\,IV and Si\,IV doublets.
Figure\,\ref{pa1} exhibits the variation of these lines during epochs 1979.5
and 1979.8. Although not plotted, the Si\,III $\lambda$1206 line shows the
same enhanced absorption out to a common edge of $\sim$-1300 km\,s$^{-1}$.
In Fig.\,\ref{pa1}
the blue wing of the He\,II line is enhanced out to  -500 km\,s$^{-1}.$
Likewise, Figure\,\ref{pa2} exhibits both a blue wing strengthening
red wing weakening in 1993.9 relative to 1979.5.
The Si\,III and Si\,IV resonance lines show similar variations as C\,IV
over the intervals covered in our two figures. Because the statistical tool
we described in $\S$\ref{statan} evaluates changes for a chosen wavelength 
interval, we exhibit the significance of our test for just the
low negative velocities of He\,II defined by the comb symbol.

\vspace*{-.15in}

\subsubsection{2 Vulpeculae } %   - c4he2 or he2}

   Although 2\,Vul has not been classified as a Be star, we
included it in our sample because Zaal et al. (1997) had detected emission
in its near-infrared Brackett hydrogen lines. This
emission implies the presence of a thin disk.  Although Percy et al. 
(1988) have reported that this star has a photometric period near 0.61 days,
Balona (1995) found a period of 1.27 days. Hanula \& Gies (1994)
discovered regular line profile variations, suggesting that these
variations are due to nonradial pulsation.  Prinja (1989) has determined
a mass loss rate of 1$\times$10$^{-9}$ $M_{\odot}$ yr$^{-1}$ from
resonance lines of several ions.

  The {\it IUE} archive includes 39 {\it SWP} observations, of which 
19 were recorded in a campaign at 1992.7. GBS and
ten Hulve (2004) found that the C\,IV lines are variable. Ten Hulve
considered it a ``magnetic star." The mean C\,IV profiles are extremely
strong, indeed so much so that the two components merge into a 
single broad trough with a minimum at $\sim$-400 
km\,s$^{-1}$. Occasionally, the small-scale variations take the form of 
narrow {\it emissions} centered at a few high velocities along the 
profile. In this sense the resonance lines are more characteristic of an 
active O star than a Be star. 

\begin{figure}
 \centering
 \includegraphics[width=6.5cm,angle=90]{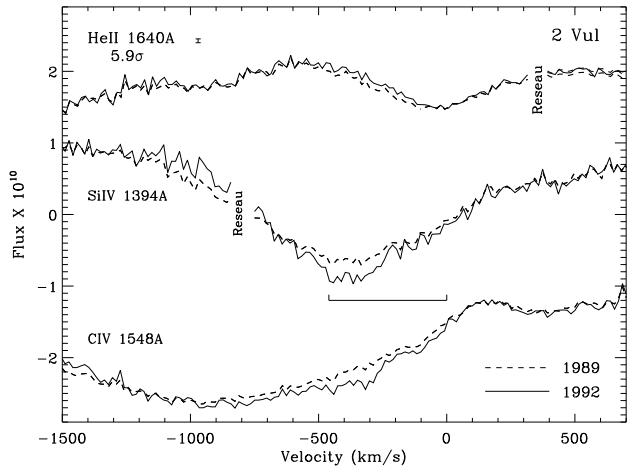}
%FIGURE 20 -  (figure 22)
 \caption{A comparison of the He\,II, C\,IV, and Si\,IV $\lambda$1394
 line spectra between epochs 1989 (SWP\,36329) and 1992 for 2\,Vul.
 The comb shows the region of differences of the profiles, which are
 also shared by the Si\,III $\lambda$1206 line (not shown).
  }
 \label{2v8992}
 \end{figure}

Rapid He\,II variations are not discernible in the {\it IUE} spectra of
this star.  We have selected two examples of 
long-term variability.  Figure\,\ref{2v8992} shows that a difference
in the profiles of He\,II, the C\,IV doublet, and Si\,IV $\lambda$1394
between an observation in 1989 and the epochal average for 1992. It
is clear that the wind was far stronger during the latter epoch and
produced enhanced absorption out to -1200 km\,s$^{-1}$.
The behavior of the Si\,III $\lambda$1206 line, not shown, is similar
to Si\,IV. This fact suggests that the variations are due to 
large increases in mass rather than a shift in wind ionization. 
In contrast, the DAC in the Si\,III profile is centered only at -800  
km\,s$^{-1}$, indicating that the position of this feature is governed by 
wind ionization. The blue wing of the He\,II line is uniformly stronger in 
1989 than 1992 and extends to -400 km\,s$^{-1}$.

\begin{figure}
 \centering
 \includegraphics[width=6.5cm,angle=90]{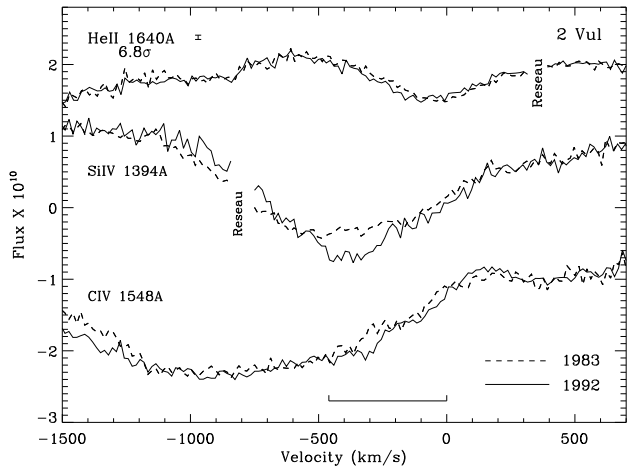}
%FIGURE 21 -  (figure 20)
 \caption{A comparison of the He\,II, C\,IV, and Si\,IV $\lambda$1394
 spectra for 2\,Vul for the epochs 1983 and 1992. Although not shown,
 the Si\,III line exhibits similar variations. }
 \label{2v8392}
 \end{figure}

  Our second example of $\lambda$1640 variability 
is shown in Figure\,\ref{2v8392}. This figure compares the behavior of 
He\,II, C\,IV, and Si\,IV lines between 1983 and 1992. This contrast is 
smaller than the previous one with respect to 1989. 
The C\,IV, Si\,IV, and Si\,III (not shown)
lines show variations in an opposite sense from He\,II --
for example, in the range -700 to -1000 km\,s$^{-1}$. This suggests
that ionization shifts are at play in this case. The strengthening
of the blue wing of the He\,II line is small but consistent out to
at least~~ -500 km\,s$^{-1}$. Although this slow change occurs along
the photospheric profile too, this is another case in which the
blue-shifted absorption is formed in the wind.

\vspace*{-.15in}

\subsubsection{19 Monocerotis } %  \#2, 1}

  19\,Mon, the final star for which we found $\lambda$1640 variability,
is also a rapidly rotating B1e star near the main sequence. It 
exhibits at least two large-amplitude prograde nonradial pulsations with 
periods near 5 hours (Balona et al. 2002).  Although the star's
Be type is based on emission twice detected at low dispersion, Balona et al.
have disputed whether the H$\alpha$  line really has ever displayed emission. 
However, GBS and ten Hulve 
(2004) have noted the variations of its C\,IV lines.
% among the 15 available {\it IUE/SWP} echellograms. 
Ten Hulve considered this another example of a ``magnetic star." 
For this reason, we are inclined to believe the original spectroscopic 
emission reports and have included 19\,Mon in our Be star sample.

\begin{figure}
 \centering
 \includegraphics[width=6.5cm,angle=90]{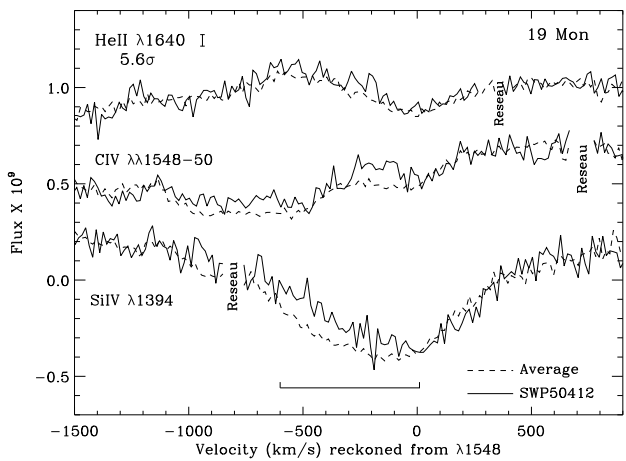}
%FIGURE 22 -  (old figure 24)
 \caption{A comparison of the He\,II, C\,IV, and Si\,IV $\lambda$1394
 spectra for a mean and an observation of 19\,Mon during 1994.2. 
 The significance on He\,II refers to the {\it whole profile.} Thus, the significance
 of the variation of the blue wing exceeds 6.0$\sigma$.
}
 \label{19mions}
 \end{figure}

  Figure\,\ref{19mions} exhibits the only possible example of a likely 
deviation of the He\,II line from its mean profile from the 15 available 
{\it IUE} spectra. The diminished absorption in this observation 
(SWP\,50412) begins at line center and extends out to
-700 km\,s$^{-1}$. Remarkably, this variation is 
anticorrelated with the C\,IV and Si\,IV strengthenings. Yet, it appears
once again that the He\,II formation region extends into the wind. 

% \begin{figure}
%  \centering
% \includegraphics[width=6.5cm,angle=90]{f25.eps}
% %GIGURE 25 -
%  \caption{19 Mon: A montage of five He\,II spectra of 19\,Mon
%   against the mean spectrum.
%   }
% \label{19vm}
%  \end{figure}

%  Figure\,\ref{19vm} exhibits five $\lambda$1640 observations for which
% the individual profiles seem to show fluctuations from the average profile. 
% We have no practical way of confirming that these fluctuations are genuine 
% in every case. However, we note the pattern that three of the observations 
% exhibit a short-lived absorption. These fluctuations
% resemble those we reported for $\lambda$\,Eri and $\omega$\,Ori. 
% A probable red emission spike is displayed in the top example of this
% figure. It appears similar to the spike discussed in Fig.\ref{hd675v} 
% in HD\,67536. 

\section{Discussion}
\label{discc}

\subsection{Characterization of the variability}

   The He\,II $\lambda$1640 variations we have found fall into one
of the following patterns:

\begin{itemize}

\item The line preferentially fills in on the red side 
    ($\lambda$\,Eri, $\mu$\,Cen,  6\,Cep, and $\eta$\,Cen; e.g., 
    Figs.\,\ref{le9395}, \ref{le1}, \ref{mc1}, 
%\ref{mc2}, 
    \ref{6c1}, \& 
    \ref{ec4}). These events are almost always also observed in the C\,IV 
    doublet and often in Si\,IV.
  
\item The whole profile fills in, or ``weakens" ($\lambda$\,Eri,
    $\omega$\,Ori, and $\mu$\,Cen; e.g., Figs.\,\ref{le2}, \ref{le5},
    \ref{oo83}, \& \ref{mc3}).  This activity is largest 
    in C\,IV and is occasionally visible in Si\,IV.

\item Discrete absorptions can occur in the middle of the profile
    ($\lambda$\,Eri, $\omega$\,Ori, $\psi^1$\,Ori, and $\pi$\,Aqr; 
    e.g., Figs.\,\ref{le2}, \ref{oo5}, \ref{po1}, \& \ref{pa1}). 
%    \& \ref{19vm}). 
    Except for the 19\,Mon events, these absorptions 
    were also found in the C\,IV doublet.

\item Short-lived, narrow emission spikes can appear in the profile
 (HD\,67536; e.g., Fig.\,\ref{hd675v}). The C\,IV doublet appears to
 respond to these events, at least for the events observed in HD\,67536.

\item Extended absorption in the blue wing of $\lambda$1640 develops
    at certain times in half the stars in our sample. Thus, this is
    the most common type of variability. The enhanced absorption
    is correlated with extended high velocity wind absorption in
    the C\,IV and/or Si\,IV doublets ($\psi^1$\,Ori, $\eta$\,Cen,
    2\,Vul; e.g., Figs\,\ref{po2}, 
%   \ref{ec2}, 
    \ref{ec3}, \& \ref{2v8992}).

\end{itemize}

\subsection{What atmospheric changes produce the $\lambda$1640 variations?}
\label{atmchg}

   Most of the morphologies just summarized involve similar variations in 
the C\,IV and/or Si\,IV doublets. Therefore, they probably reflect changes in 
the structures of the stars' winds. For the spectra of $\lambda$\,Eri at least,
we also note that some of these events correlate
with the appearance of He\,I optical line dimples -- that is, 
a sympathetic response in the C\,IV lines can also be found (Smith et 
al. 1996).  We have stipulated that there are some He\,II events for which 
correlations with extant optical or UV data do not occur. We cannot
conclude much from these particular cases.

  A common event class, second in overall frequency to those displaying
an extended blue wing type, is the filling of the  central and red-wing
profile regions, and sometimes the whole line. Attempts to interpret such 
events encounter the ambiguity of weakened absorption versus true emission,
and thus cannot be straightforwardly attributed to changes in the wind.
For example, weakened absorption might be caused by a decreased effective 
temperature over the visible disk, or by a decrease of the temperature 
gradient within the photosphere.  Were either a lowering of the $T_{\rm eff}$
or the temperature gradient to occur for whatever reason, it would have the
effect of decreasing the equivalent widths of all excited photospheric lines, 
as well as the wind components of the lines we have studied.
This prompts the question of whether the general
weakenings of the He\,II line also correspond to less blue wing absorption 
in the C\,IV lines. In the examples we have shown with general line 
weakenings, the answer seems to be ``yes" in about ${\frac 23}$ of the cases. 
This includes the events shown for $\eta$\,Cen, 2\,Vul, 19\,Mon, 
and the Fig.\,\ref{pa1} event for $\pi$\,Aqr. Thus, it 
appears that an argument can be made that the thermal conditions of the 
photosphere have a effect on the velocity and density 
relations of the wind.

  Smith et al. (1997) have constructed non-LTE models of static atmospheres
in order to examine the requirements needed to reproduce observed emissions 
(or absorption weakenings) in $\lambda$1640 and the red He\,I lines 
of $\lambda$\,Eri. 
They found that emission can be produced within a moderately dense, heated
slab above the atmosphere by ``Lyman pumped recombination." In this
process the slab's helium atoms feel the effects of the
slab's own Lyman continuum radiation. He\,I line emission will result if the 
slab is heated to about 50,000K or illuminated by EUV flux having an 
equivalent radiation temperature. The process is efficient
%, first, 
for a slab density of $\sim$10$^{11-12}$ cm$^{-3}$ (which is incidentally the
typical density where the line cores of the He\,II line are formed in B
dwarf atmospheres). Moreover, the slab should be thick enough for the 
helium lines to be optically thick, while at same time allowing the 
Lyman continuum to remain thin.  Such emitting structures might take 
the form of mild density blobs suspended above the photosphere.
Equivalent conditions might be produced 
by a flattening of the density lapse rate in its upper regions. 

  What causes the filling in of the red wing of the He\,II line?
According to VCR, a heated, isotropic 
wind, even with the mass loss rate expected for an early Be star, is capable
of producing visible P\,Cygni signatures in the He\,II, Si\,IV, and
C\,IV lines. However, except in Fig.\,\ref{hd675v} (HD\,67536) the incipient 
redshift emissions we have found are {\it not} accompanied by blueshiftings 
of the absorption components: when the red wing is raised, the blue half
of the profile is usually unchanged. This observation runs contrary to
the predictions of P\,Cygni profiles for $\lambda$1640 and $\lambda$4686 
from VCR models; the models predicting these features included
strong, fast winds (\.M = $\ge$ 10$^{-7}$ M$_\odot$ yr$^{-1}$) and  
an equivalent $\beta$ $<<$ 1) heated to about 10,000\,K above 
$T_{\rm eff}$; see also Hamann \& Schmutz (1987).
Under these conditions line emission is produced efficiently because 
recombinations to He$^{1+}$ are sensitive to high density and 
temperature. If the region coincides with
the base of the wind, the acceleration of the flow reduces flux shielding
in the wind, permitting more atoms to be exposed to the deep photospheric
radiation flow.  In $\S$\ref{reslt} we emphasized that the red emissions
of $\lambda$1640 should not be described as true P\,Cygni profile.
The VCR models of $\lambda$1640 indicate that P\,Cyg profiles are most
easily produced if a heated chromosphere exists very close to a star.

  In two of the examples we discussed, Figs.\,\ref{mc3} and \ref{6c1}, the 
response of the C\,IV doublet to weak red wing emissions in $\lambda$1640 is 
accompanied by increased blue wing absorption. 
This fits with VCR's modeling results that a heated region can be placed 
at a position too far from the star to influence the He\,II line but yet 
where it would be still responsible for C\,IV absorption in the wind.
It is also important to point out that because VCR's ``distant, heated 
wind slab" models produce red wing $\lambda$1640 emission, one does not
have to resort to {\it ad hoc} ``returning blobs" to explain this 
emission in the observed profiles.
 
  Before undertaking their work on the He\,II line,  Venero, Cidale,
\& Ringuelet (2000) had produced anomalous wind models that led to 
absorption and emission signatures in the C\,IV doublet
similar to those they found later for He\,II.
Although similar models for C\,IV have not been explored yet by any
authors, one can surmise that the responses of these lines would be 
similar to those just outlined. One might guess that the effects on 
C\,IV would be amplified for those models in which the wind is heated 
far from the star. For example, weak
red wing emissions are most visible in the C\,IV observations 
-- see Figs.\,\ref{mc1}, \ref{6c1}, and arguably \ref{hd675v}.  
The mildness of this emission may be used in the future models to  
constrain the distance of heated regions above the star.

  In addition to the wind heating requirements, the work of Venero, 
Cidale, \& Ringuelet demonstrates the intuitive result that variable
He\,II characteristics, whether in absorption or emission, 
increase with the mass loss rate. 
In cases where we have found correlated blue wing variations in 
both Si\,III and Si\,IV lines (e.g., Figs.\,\ref{pa1} and \ref{2v8992}) 
we estimate from tests of moving slab models using the {\it CIRCUS}
program (Hubeny \& Heap 1996)
 that the mass loss rate must be enhanced by a factor of at least
a factor of 10.  This enhancement is too large to be an 
effect of refocusing of wind in a magnetic dipolar field.

  We have argued that
the high velocity absorptions of the He\,II line can be best 
understood by a change in the mass loss rate and probably the velocity
acceleration law. In addition, the most likely explanation for the faint 
red wing emissions is that an unknown instability, possibly magnetic, 
heats an accelerating region of the wind. 
Finally, we have suggested that conditions within the photosphere are 
responsible for the relatively common line weakenings across the 
photospheric profile.

\section{Conclusions}

  This paper provides a mini-atlas of He\,II $\lambda$1640
variability for a group of 10 Be stars selected from 
a much larger sample of early-type Be, Bn, and B normal stars.
We have identified several basic types of variability. Weak red emissions 
and line weakenings occur over timescales of a few hours or less. 
In terms of the variability timescales, we have noted that
the pattern of strengthening blue wings occurs over long timescales.
In our view this is most likely explained by changes in the wind
velocity law (cause unknown). Second, line weakenings
likewise occur preferentially over long timescales. Long-term 
weakenings occur in half of our ten $\lambda$1640-variable stars. 
Third, weakenings over the whole line or only the red wing can occur 
even within a few hours. We also point out that rapid variability was
found preferentially in the stars $\eta$\,Cen and $\lambda$\,Eri. 
We believe these events speak to intrinsic properties of these stars rather
than to observational sampling.

   We have suggested that the properties which
change the surface and wind properties of these stars are mediated 
by magnetic instabilities. This is among the few ways of 
interpreting aperiodic variations of a single star on a timescale sometimes
much less than its rotation period. We note that our examples of
variable $\lambda$1640 do not include known Bp stars. Magnetic
fields in these stars are thought to be dipolar and, most importantly,
stable over at least several years. Strong stable fields resist the influence
of hydrodynamical instabilites that might alter a  wind's structure 
or its geometrical flow. Therefore, we conjecture that at least for 
some of these stars magnetic fields must be localized on the 
surface.  Velocity perturbations due to nonradial pulsations and
differential surface rotation would then offer plausible ways to 
trigger magnetic instabilities in these multipolar configurations.

\acknowledgements{ We thank Dr. Roberto Venero for clarifications on 
his work on the He\,II $\lambda$1640 line, Dr. Geraldine 
Peters for making available H$\alpha$ observations of $\omega$\,Ori, and
an anonymous referee for improving the quality of this paper.
}

\end{document}